\documentclass{article}

\title{\textbf{Assessing the impact of weather-induced uncertainties in large-scale electricity systems}}
\author{Jan Peper, David Kröger, Jonathan Kipp, Florian Ziel, Christian Rehtanz}
\date{}

\usepackage[a4paper, left=3.5cm, right=4.5cm, top=3cm]{geometry} 
\usepackage[utf8]{inputenc}
\usepackage[T1]{fontenc}
\usepackage{lmodern}
\usepackage{microtype}
\usepackage{amsmath,amssymb,amstext}
\usepackage{siunitx}
\usepackage{upgreek}
\usepackage{graphicx}
\usepackage{color}
\usepackage{xcolor}
\usepackage{booktabs}
\usepackage{multirow}
\usepackage{float}
\usepackage{subcaption}
\usepackage[backend=bibtex,style=numeric,sorting=none]{biblatex} 
\usepackage[printonlyused]{acronym}
\usepackage[pdftex,hidelinks]{hyperref}
\usepackage[nottoc]{tocbibind}
\usepackage{lineno}
\usepackage{url}
\usepackage{xurl}

\begin{document}
	
	\maketitle
	
\begin{abstract}
	
	\textbf{The future energy system will largely depend on volatile renewable energy sources and temperature-dependent loads, which makes the weather a central influencing factor. This article presents a novel approach for simulating weather scenarios for robust large-scale power system analysis. By applying different signal analysis methods, historical weather data is decomposed into its spectral components, processed appropriately, and then used to generate random, self-consistent weather data. In this process, any weather parameters of different locations can be considered, while their respective dependencies are mapped. The added value is demonstrated by coupling with a state-of-the-art large-scale energy system model for Europe. It is shown that the integrated consideration of different weather influences allows a quantification of the range of fluctuation of various parameters - such as the feed-in of wind and solar power - and thus provides the basis for future resilient grid planning approaches. }	
	
\end{abstract}
    
The European energy system is currently in a transition phase, with meteorological conditions increasingly influencing the behavior of various actors. On the supply side, the weather mainly affects the generation of renewable energy sources (RES) \cite{weather_on_ee, weather_on_ee2}; on the demand side, especially the temperature has an influence on the operation of weather-dependent loads like electric heat pumps (HPs) or electric vehicles (EVs) \cite{ETG_David, BEVTemperaturabhengigkeit}, and consecutively on prices \cite{ZIEL2018251}. Conventional approaches to electricity market modelling usually represent the metrological influence of a single or few selected historical weather years \cite{TYNDP2020} which raises doubts about the resilience of existing and future energy infrastructure in the context of an increasingly volatile feed-in with flexible weather-dependent consumers \cite{WeatherYears}. This is also recognized by the \textit{European Network of Transmission System Operators for Electricity} (ENTSO-E), which plans to include a more significant amount of historical weather data in their future network development plans \cite{ENTSOE_Wetter}. Nevertheless, since the availability of consistent historical weather data is limited, its exclusive use for long-term energy system planning seems problematic, as the range of potential future weather events may not be fully represented, and no reliable information about the underlying distribution of the behaviour of different actors can be obtained \cite{ClimateRiskAssessment}. 

To map these weather-induced uncertainties, a novel spatio-temporal simulation model based on a Fourier analysis \cite{fourier_2009} that considers the influences of natural weather fluctuations at different time scales and locations is proposed. This model is coupled with the established large-scale energy system model \textit{MILES} \cite{NEMO}, allowing the quantification of the effects of various weather influences on the operation of RES, temperature-dependent loads and the dispatch of European power plants. 

The central approach is the generation of random weather data, which can depict possible future events. In the context of large-scale energy system analysis, this leads to the challenge of generating consistent time series in hourly resolution, taking into account dependencies between both locations and parameters. 

Two approaches to generate time series in accordance with above-mentioned requirements lie on the one hand in the use of physical weather models which incorporate the underlying fundamental relationships between weather parameters \cite{guillod_weather_at_home, eggimann_weather_heating} or on the other hand by means of statistical time series analysis \cite{nature_extreme, knight, weather_building}. However, realistic energy system modeling requires input data with a consistent spatio-temporal dependency structure \cite{spiekerDiss}, which is typically lacking due to the focus on a single location or simplified assumptions leading to the fact that "the time series of the expected scenarios do not represent realistic variations of the climate system" \cite{nature_extreme}.

In general, the consistent modeling of weather data for large-scale energy system analysis that takes into account both spatial and temporal correlations between individual time steps and parameters in the time domain is considered a significant challenge \cite{timecoupling}. To address this challenge, this article presents a model based on a transformation of the weather data to the frequency domain, which maps the time series as the sum of individual oscillations. This approach is in line with the modelling proposed by the \textit{Intergovernmental Panel on Climate Change} (IPCC), which represents the weather as a sum of trends and seasonal or random fluctuations \cite{IPCC}. The analysis of data in the frequency domain has already been investigated in different contexts such as periodicity of temperature at a single location \cite{Bialystok}, interpolation of missing data \cite{weather_year_china}, load forecasting \cite{Demand_Turkey} or the clothing industry \cite{fashion}, although no application to the generation of large-scale weather data for energy system analysis has been accomplished yet.

\section*{Modelling weather-induced uncertainties}
\begin{figure}[ht!]
	\centering
	\includegraphics[width=132mm]{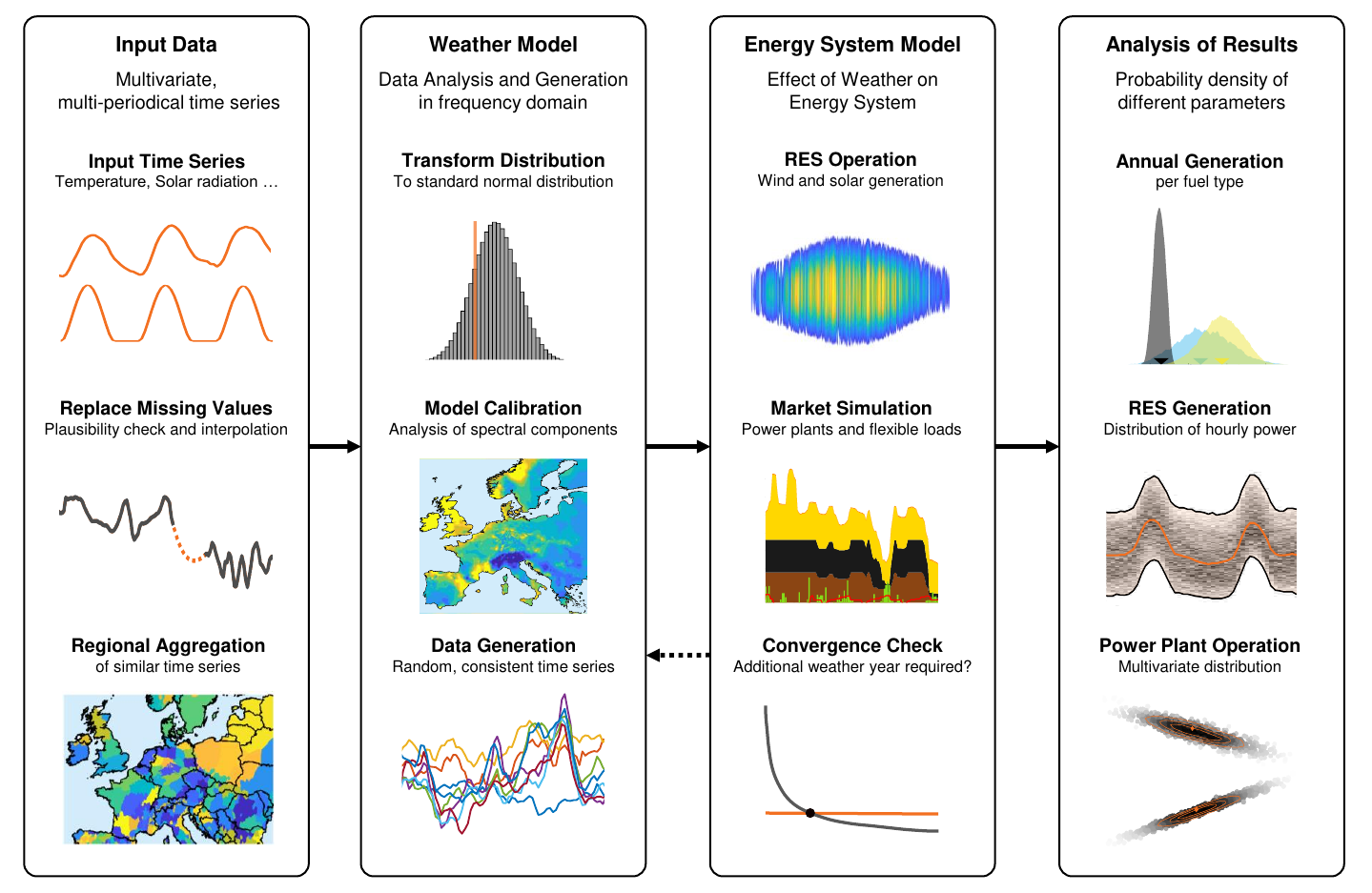}
	\caption{Overview of the proposed model. Multiple time series of weather parameters of different locations are imported and preprocessed. This is followed by a regional aggregation of similar time series. In the \textit{Weather Model}, a transformation to the frequency domain is carried out, which is used to calibrate the model and to generate random time series. These are used in the \textit{Energy System Model}, which includes a convergence check to obtain the required amount of random weather years. The results comprise the effect of the weather on the distribution of all parameters that are considered within the energy system model, such as RES generation and power plant operation.}
	\label{fig:Overview}
\end{figure}

First, the different steps of the analysis and generation of random weather data are described and subsequently the results are presented. We refer to the \textit{Methods} section for a detailed mathematical description. Figure \ref{fig:Overview} gives a overview of the different steps of data analysis and generation as well as the coupling to the energy system model. 

\subsection*{Concept}
The basic approach is the decomposition of the weather time series into their spectral components using the Fourier transformation. This allows the data to be mapped as a sum of superimposed oscillations with different period durations, amplitudes and phases. These individual oscillations can be interpreted as different influences of different weather phenomena; for example, an oscillation with a period of one year and a large amplitude represents the seasonal fluctuation, while superimposed oscillations can be related to different random weather effects. Based on a statistical analysis of these different spectral components, an arbitrary number of random weather years can be generated, preserving both the seasonal trend and different weather influences in various time ranges. 

\begin{figure}
	\centering
	\includegraphics[width=135mm]{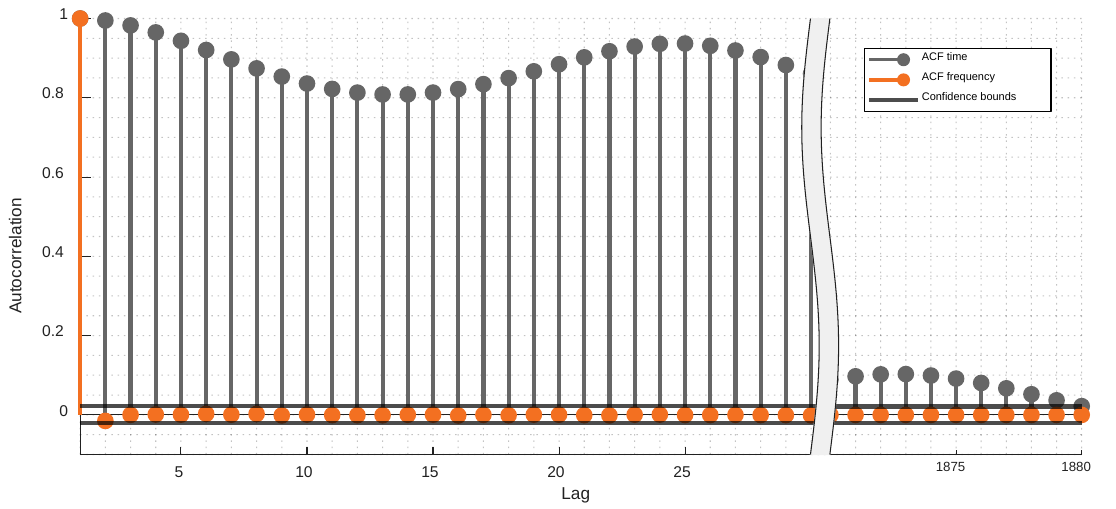}
	\caption{Autocorrelation of temperature data in the time and frequency domain with 95\,\% confidence bounds. The correlation of neighbouring time steps is evident over a long period of time, while the dependence of adjacent frequencies is significantly lower, showing the benefit of analysis in the frequency domain. } 
	\label{fig:akf}
\end{figure}

While an analysis of weather data in the time domain requires the consideration of dependencies between multiple time steps, switching to the frequency domain allows the independent analysis of individual spectral components, which leads to a considerable reduction in the model complexity (Figure \ref{fig:akf}). 

\subsection*{Input Data}
In principle, different types of data can be used as input for the random data generation. A consistent data set is required, in which periodicity and correlations between different data series are mapped. In this article, hourly weather data of temperature, wind speed and global solar radiation of 21 years for various European locations as provided by the German Meteorological Service \textit{Deutscher Wetterdienst} (DWD) \cite{DWD_REA6} is used. 

\subsection*{Preprocessing}
Before the analysis of the data with respect to their fundamental properties and correlations takes place, the data is modified in such a way that their distribution matches a standard normal distribution. This ensures that different types of input data can be treated independently of the underlying properties of the respective parameter while their specific characteristics and temporal patterns are preserved. The normalization is achieved by computing the empirical distribution function of each input time series. By using the inverse distribution function, the values of each time series are modified to follow a standard normal distribution. The empirical distribution functions are stored separately for each time series and are later used to denormalise the generated data.

In addition, because solar radiation must always be zero at night, it requires pre- and post-processing in order to be appropriately used with the described method. In the first step, all time steps with no possible solar radiation due to the time of day are identified and classified as \textit{night}. From a signal analysis point of view, these time steps can be seen as missing data, since it is a gap in the time series leading to spectral leakage as described in \cite{MIARMA}, which would harm the analysis. In order to diminish this effect, the zero values during night-time are removed from the input data, and replaced by interpolated values which fictitiously continue the solar irradiation at night by means of an autoregressive fits, as described in \cite{autoregressive}. Another characteristic of solar radiation data is that the random deviation of the current radiation from the long-term average depends on the current radiation intensity. To account for this issue, the solar data is normalized by the long-term hourly mean of all years considered per location so that only the normalized deviation from the long-term mean is used as input data for the generation of new radiation data. After the data generation, the resulting data are de-normalized using the same long-term hourly mean value for each location.

\subsection*{Analysis in Frequency Domain}
In this step, the normally distributed weather time series are transformed into the frequency domain separately for each year and parameter using the Discrete Fourier Transform (DFT). 

\begin{figure}
	\centering
	\includegraphics[width=135mm]{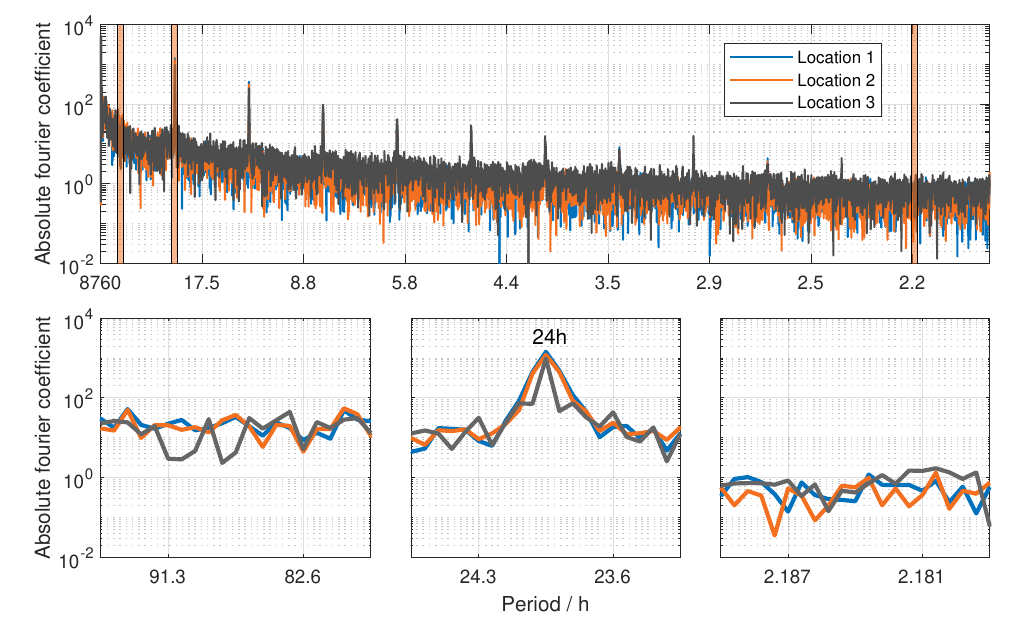}
	\caption{Absolute value of Fourier coefficients of three locations. The correlation between close locations 1 and 2 is higher compared to remote location 3. Furthermore, the correlation and the intensity decrease with a lower period duration. The peak at a duration of 24 h is an effect of the daily fluctuation, which is present in all time series.}
	\label{fig:verlauf_korrelation}
\end{figure}

The correlation between different locations and parameters is calculated based on the obtained complex Fourier coefficients\footnote{A test for normal distribution of individual coefficients is carried out.}. This is done separately for each frequency, since the correlation can be influenced by varying factors in different frequency ranges. In general, it can be observed that the correlation of the Fourier coefficients of corresponding oscillations decreases with increasing frequency and distance between the locations. This suggests that long-term weather phenomena usually affect larger regions, while fluctuations in shorter intervals have a more regional effect (Figure \ref{fig:verlauf_korrelation}). 

In addition to the absolute values, the phase angle of all oscillations is also analysed. Figure \ref{fig:phase} shows the phase angle of the Fourier coefficients at the frequency representing the daily fluctuations. It is evident that the western locations tend to have lower phase angles, which can be explained by the later time of sunrise. Since the phase angle is obtained for each location in the process of data analysis, these specific characteristics will also be present in the generated data.  

\subsection*{Data Generation in Frequency Domain}
This step aims to generate random complex Fourier coefficients for each frequency, which on average, follow the previously determined correlations and mean values. A consistent set of correlated coefficients is obtained for each generated weather year.

The amount of generated random years can be defined as desired or obtained by the convergence of selected parameters. In general, determining the number of required weather years is a critical problem - too few may not adequately represent all weather phenomena - too many will lead to an unnecessarily high computational effort.  

\begin{figure}[ht!] 
	\centering
	\begin{subfigure}{0.49\textwidth}
		\includegraphics[width=\textwidth]{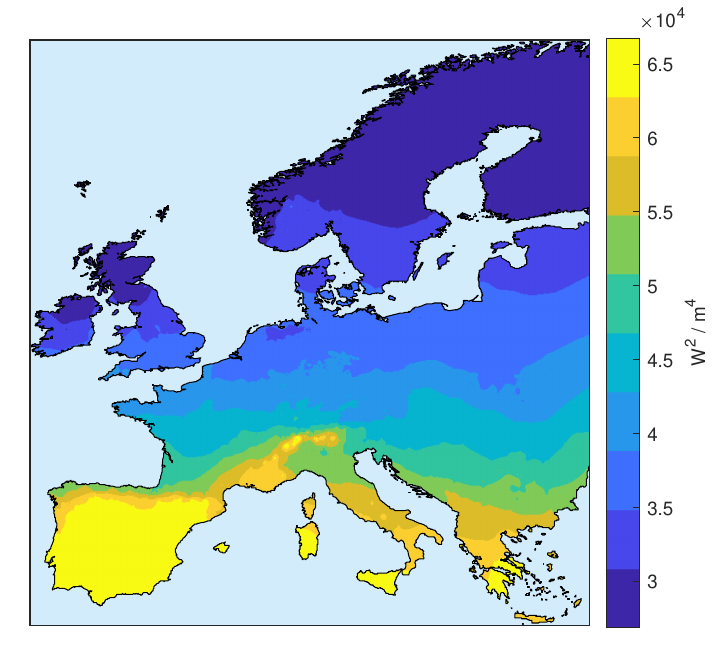}
		\caption{Regional distribution of noise intensity \\ of global solar radiation}
		\label{fig:noise_solar}
	\end{subfigure}
	\hfill
	\begin{subfigure}{0.48\textwidth}
		\includegraphics[width=\textwidth]{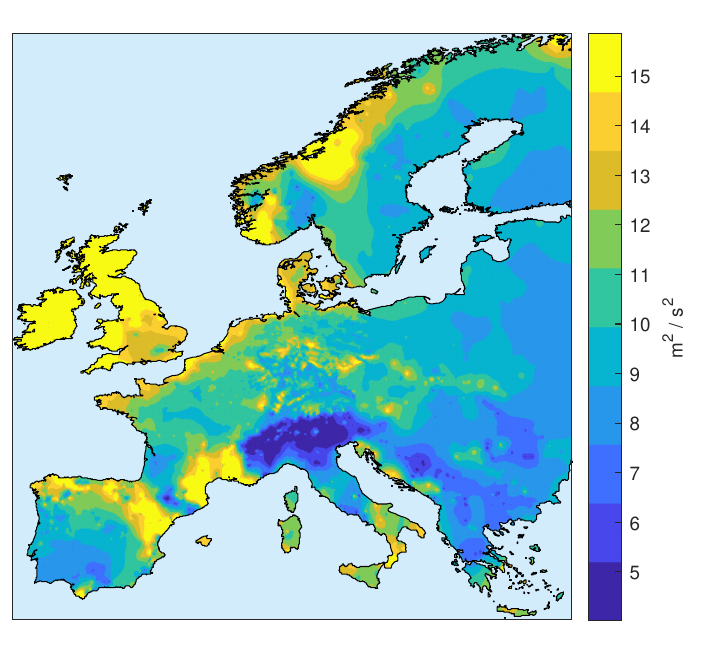}
		\caption{Regional distribution of noise intensity \\ of wind speed}
		\label{fig:noise_wind}
	\end{subfigure}
	\begin{subfigure}{0.49\textwidth}
		\includegraphics[width=\textwidth]{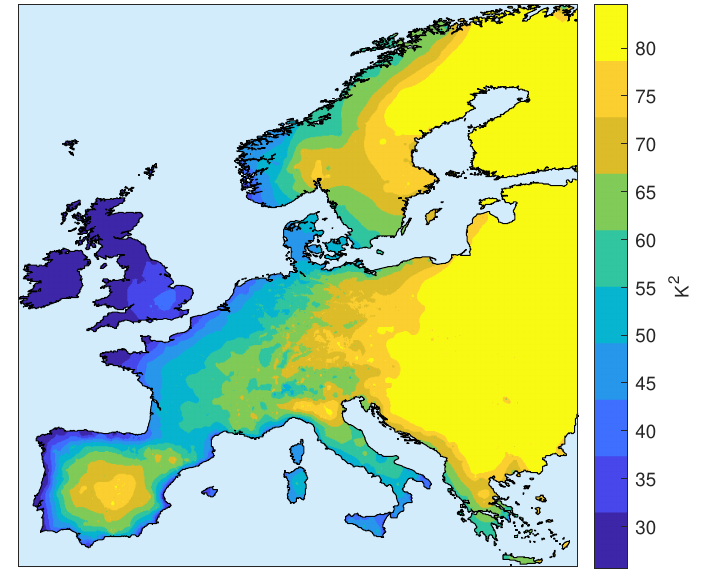}
		\caption{Regional distribution of noise intensity \\ of ground temperature}
		\label{fig:noise_temperature}
	\end{subfigure}
	\begin{subfigure}{0.48\textwidth}
		\includegraphics[width=\textwidth]{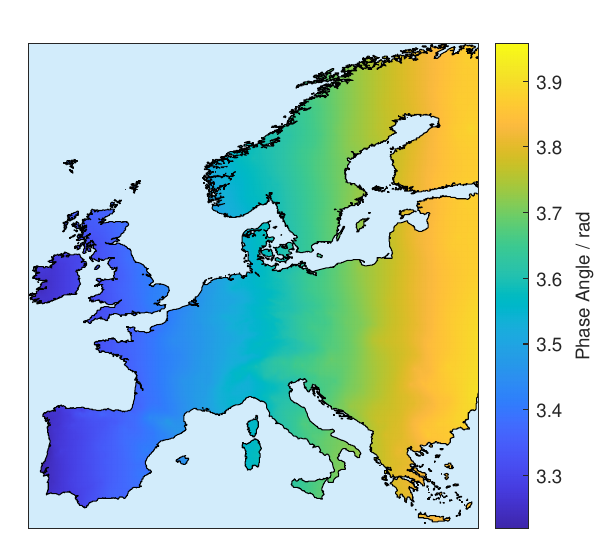}
		\caption{Phase angle of daily fluctuations \\ of solar radiation.}
		\label{fig:phase}
	\end{subfigure}
	\caption{European noise intensity of different weather data. Regions with higher intensity tend to have higher fluctuations and are, therefore, more affected by increased future weather uncertainties. Furthermore, the Fourier transformation determines a plausible relationship between local phase angles, which represent the different times of sunrise.}
	\label{fig:noise}
\end{figure}

An essential part of generating new weather data is the proper representation of the intensity of different oscillations, which can be interpreted as noise in the context of signal analysis. In this context, Figure \ref{fig:noise} shows the absolute noise intensity of the simulated time series of wind speed, global solar radiation and temperature. It is evident that the absolute fluctuation of solar radiation tends to decrease with increasing latitude, while in the case of wind generation, particularly high uncertainties can be observed in coastal regions. This noise intensity is reflected in the absolute value of the corresponding Fourier coefficients showing its proper mapping within the proposed model. 

\subsection*{Post-processing}
\begin{figure}
	\centering
	\includegraphics[width=\textwidth]{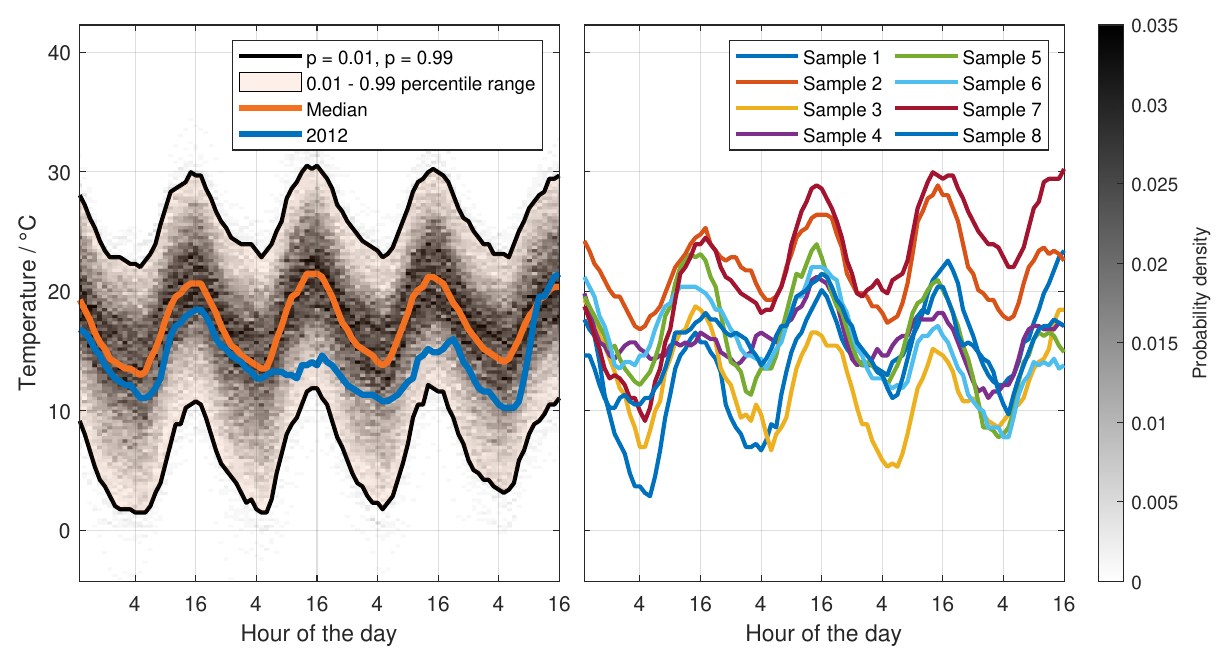}
	\caption{Left: Distribution of randomly generated hourly temperature data for an exemplary time period of 96 hours during summer. The orange area represents the 0.01 - 0.99 percentile range of all 947 randomly generated weather years to achieve convergence and the blue line the real temperature curve of the year 2012. \\ Right: Eight randomly picked generated temperature data samples of the same period. This allows the observation of the temporal dependency structure of individual trajectories, while this is not possible with the marginal distributions on the left side. } 
	\label{fig:fanchart_temperatur}
\end{figure}

Since the time series generated in the previous step are based on the normalized input time series, their values need to be transformed using the corresponding empirical distribution of the underlying input data. As a result, randomly generated time series of individual parameters are obtained, preserving the fundamental correlation between locations and parameters and their corresponding volatility. For example, Figure \ref{fig:fanchart_temperatur} shows the distribution of randomly generated temperature data for four summer days at a location in northern Germany.  

\newpage

\section*{Integration in Energy System Analysis}
Next, the coupling of the presented weather model to the existing energy system analysis model \textit{MILES} \cite{spiekerDiss, milesURL} is described and the results of the economic dispatch optimization are presented and analysed from a European perspective. Special emphasis is laid on the quantification of (several) uncertainties which are compared to the analysis of a single historical weather year. 

\subsection*{Integrated Pan-European Electricity Market Modeling}
The European interconnected power system supplies around 450 million people with an annual electricity demand of about 3660 TWh \cite{entso-e_statistical_factsheet}. In Europe, electricity market clearing is carried out based on geographically-tailored bidding zones to account for limited transfer capacities between and within member states \cite{conversations_market_design}. The economic dispatch optimization applied is a market equilibrium model which aims to match electricity supply and demand in each bidding zone taking into account possible cross-border exchange. The objective function aims to minimize the short-term system costs consisting of the sum of variable operating costs of each power plant over all time step considering specific production costs which includes the costs for primary fuel, CO\textsubscript{2}-emissions and operation and maintenance. A load coverage constraint ensures that for each bidding zone the sum of the electricity demand and net export is covered by the sum of feed-in from conventional power plants, energy storages and RES. Further constraints comprise permissible minimum and maximum power limits for power plants and energy storages as well as the maximum available capacity of storages and their state equation. 

We apply \textit{MILES} to address the following two issues. First, the number of required synthetic years to achieve convergence concerning a predefined measure of dispersion is determined. Secondly, the stochastic results and their implications for the energy system are evaluated and discussed. In order to examine the effects of weather-induced uncertainties solely, other influencing factors like installed capacities are kept constant throughout the study and are based on projections for the target year 2030. The ENTSO-E region outside Germany is modeled based on the Ten Year Network Development Plan (TYNDP) 2018 and 2020 \cite{TYNDP2018, TYNDP2020}, and the input data for Germany is based on the Federal German network development plans 2019 and 2021 \cite{TSOGermany, NEP2021}. 

We begin by generating a single synthetic weather year which serves as an input parameter for the determination of feed-in time series of PV units and wind turbines for all locations considered in Europe. Furthermore, the corresponding hourly temperatures also serve as boundary conditions for the dispatch of heat-driven conversion units like combined heat and power (CHP) plants and heat pumps. Based on these weather-dependent time series, a dispatch optimization is carried out to obtain the hourly operation of all European power plants. Next, additional random weather years are generated iteratively, and their effect on the dispatch is analyzed. If the influence af an additional weather year on the standard normal error of the power plant operation falls below a specified threshold, the calculation is terminated. In this use case, the requirement of a relative standard error of 1\,\% leads to 947 randomly generated weather years. 

\subsection*{Results: Pan-European Electricity Market Modeling}
Based on the 947 random weather years necessary to achieve convergence, the examination of results is carried out. In the following, the historical weather year 2012 is used as a benchmark because it covers typical weather conditions regarding the annual yields of RES in a year-to-year comparison in Germany \cite{TSOGermany, NEP2021} while also covering periods of severe conditions (dark doldrum\footnote{There is no common definition of the term dark doldrum. For example the TNYDP 2022 defines an adequacy case study called \textit{DunkelFlaute} which represents "a two-week cold spell with low wind load factors and solar radiation" which in turn leads to a high residual electricity demand \cite{TYNDP2022_ScenarioBuildingGuidelines}. In a year-to-year comparison of the climate years 1987 - 2016 conducted in the TYNDP 2022, 2012 ranks first in covering the most stressful \textit{DunkelFlaute} case \cite{TYNDP2022_ScenarioBuildingGuidelines}.}).

\begin{figure}
	\centering
	\includegraphics[width=\textwidth]{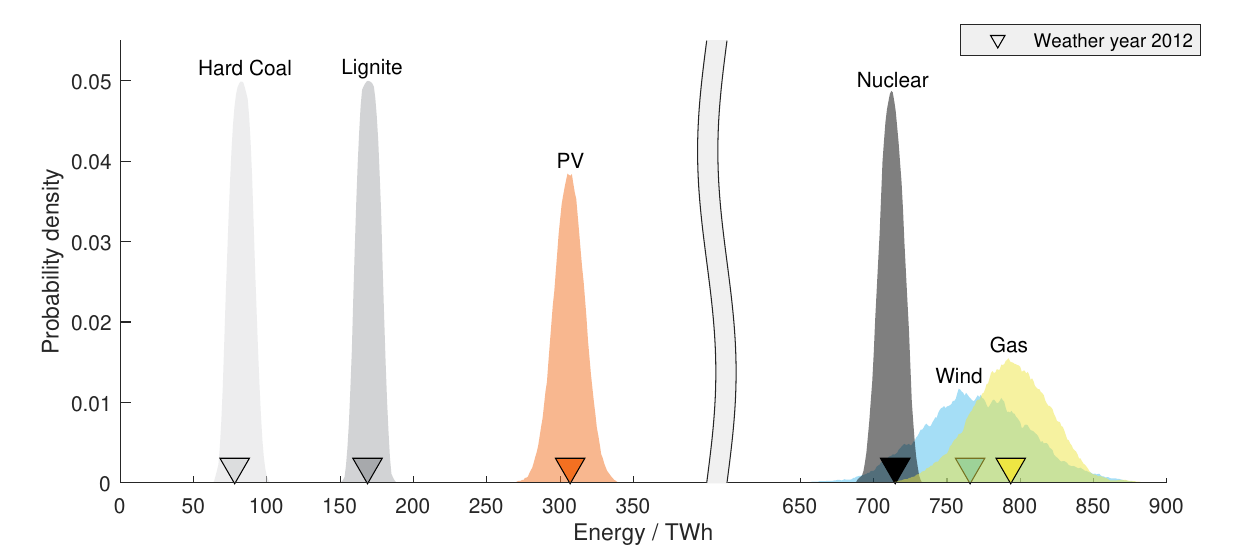}
	\caption{Probability density of annual primary fuel type generation in Europe. Wind and PV plants have the highest annual variability, which is mostly balanced by gas-fired power plants. Due to their relatively low operational costs, hard coal, lignite and nuclear power plants have high full-load hours in all weather scenarios leading to a more narrow distribution.} 
	\label{fig:AnnualEuropeanGeneration}
\end{figure}

Figure \ref{fig:AnnualEuropeanGeneration} depicts the distribution of the annual generation per primary fuel type for Europe. Generally, the distributions tend to follow a normal distribution, although for some primary fuel types, there are minor differences in the area of the mode and the tails. For example, the distribution of PV exhibits slightly larger tails except for two observations on the right side. In the case of nuclear plants, the distribution is slightly skewed to the left, with a minor plateau near the mode. 

\begin{figure}
	\centering
	\includegraphics[width=\textwidth]{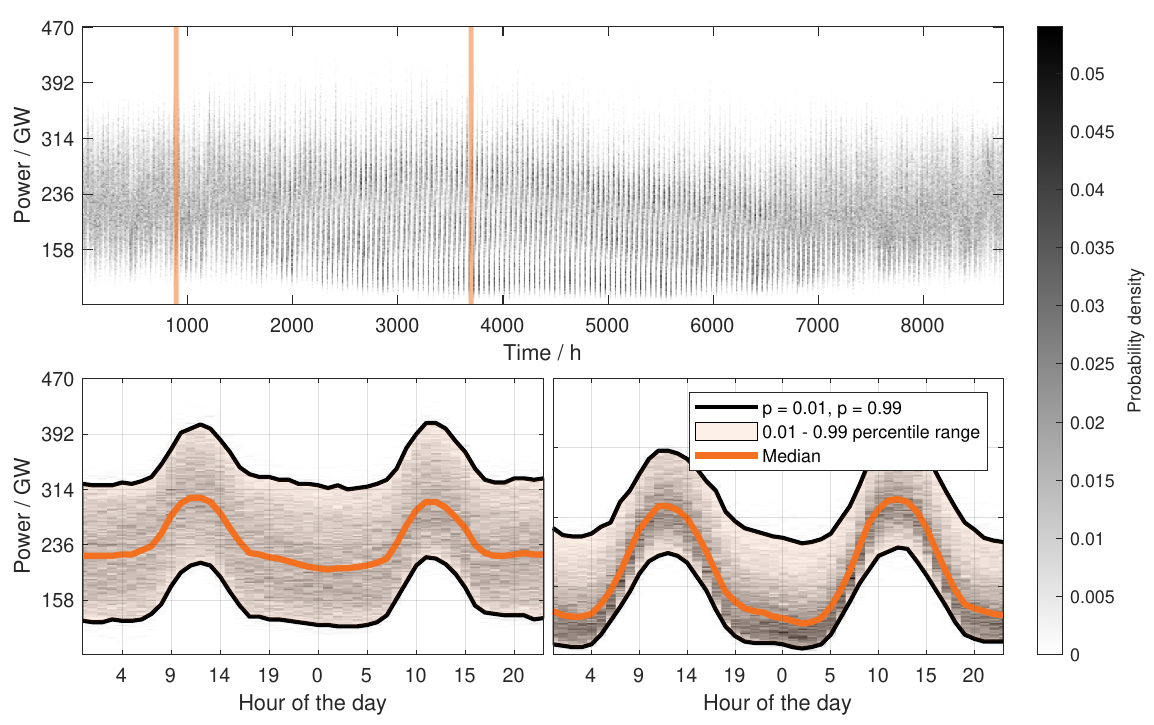}
	\caption{Probability density of the production from RES. While the upper figure shows all hours of the year, the lower figures represent selected time steps during winter and summer (marked in orange in the upper figure). The distribution of hourly generation differs between summer and winter due to the varying influence of PV and wind generation.} 
	\label{fig:RES_density_year}
\end{figure}

Time series analysis is conducted by constructing a probability density function for the operation of each hour of the year for conversion units. The production from RES exhibits a seasonal characteristic throughout the year. During winter months, feed-in from wind turbines is predominant while feed-in from PV becomes predominant in the summer months (Figure \ref{fig:RES_density_year}) resulting in a split day-to-day distribution with two focal points - one during the night hours and one during the daylight hours.

Due to having the lowest marginal cost of all power plant types considered, the production of nuclear power plants shows the highest probability density at the point of the corresponding maximum available capacity in most hours. In contrast, gas-fired power plants often change their operating points making use of their high flexibility to buffer the intermittent nature of RES. 

\begin{figure}
	\centering
	\includegraphics[width=0.95\textwidth]{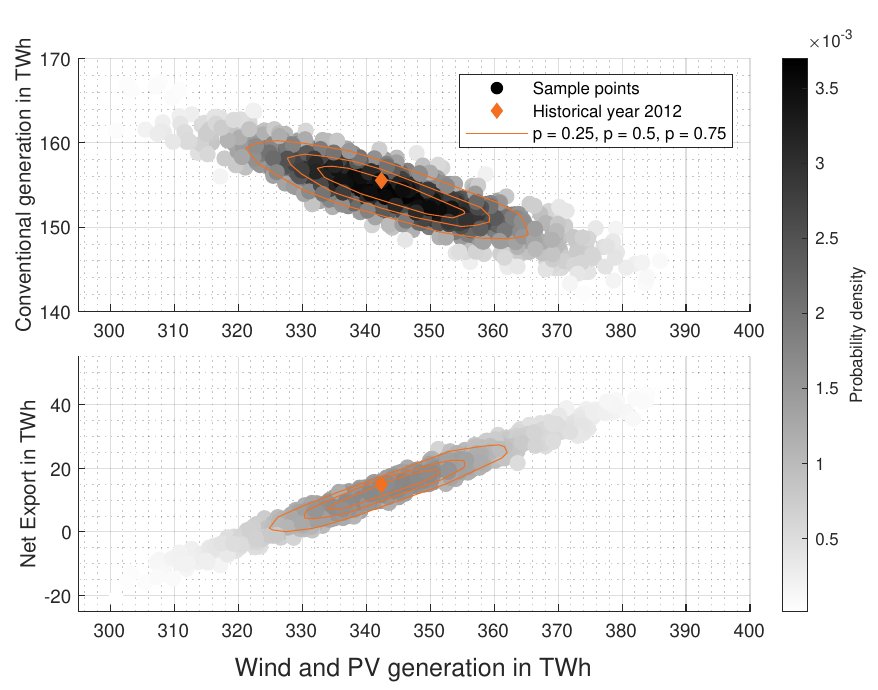}
	\caption{Probability density of wind and PV generation against (a) conventional generation and (b) net export. Orange: Historical weather year 2012} 
	\label{fig:ProbDensityGenerationExport}
\end{figure}

Next, results with a focus on the German bidding zone are examined. Figure \ref{fig:ProbDensityGenerationExport} depicts the multivariate distribution of annual conventional generation and weather-dependent RES generation as well as the dependency between annual trade balance and weather-dependent RES generation for the German bidding zone. The distribution indicates a strong correlation between annual RES and conventional generation. Furthermore, an expansion of the density cloud along the northeast to the southwest axis can be observed. This effect results from the possible trade between neighboring countries and varying annual trade balances. The distribution in Figure \ref{fig:ProbDensityGenerationExport}b exhibits a strong correlation between RES feed-in and trade balance indicating the relevance of RES generation as a driver of export. The historical weather year 2012 for both cases is located close to the centre of gravity. 

\section*{Conclusions and outlook}
We developed a novel method for synthesizing weather data for application in large-scale energy system models to enable sustainable and resilient planning. By coupling the proposed method with an existing pan-European dispatch simulation, the exposure of various parameters to weather influences is examined. The use of real scenario data in combination with high-resolution power plant data ensures a close approximation to reality. 

The transition to stochastic modelling enables an appropriate estimation of the range of possible fluctuations in, for example, RES generation or unit dispatch. We have shown that there is tremendous variability both in the hourly and total annual generation of RES, which is not adequately mapped by conventional approaches of grid planning. Even though the historical weather year 2012 with its dark doldrums, poses a suitable basis for determining weather-induced critical contingencies, the accompanying variance of parameters resulting from meteorological influences is unknown, and thus, no statement regarding confidence bounds can be made. In contrast, with our approach, significantly more robust results can be achieved compared to energy system analysis based on individual historical weather data. The results obtained by the dispatch simulation can be used in the next step as input data for a power flow calculation enabling robust identification of grid congestions under uncertainty or to asses the economic viability of different assets. The general formulation of individual weather data as a sum of superimposed oscillations ensures high compatibility with the models of the IPCC. This allows the appropriate adjustment of the intensity of individual oscillations to represent an increase in more extreme weather events due to climate change; the parametrisation of this is the subject of further research. 

In our investigation about 950 years were necessary to achieve convergence of results. This is not possible using the limited amount of historical weather years available. Generally, this leads to the question of how ideal synthetic weather years can be selected to achieve fast convergence. In a next step, it could also be investigated to what extent alternative starting and end points of the annual intervals lead to more robust results. For the present analyses, the 1$^{st}$ January and 31$^{st}$ December were chosen as limits for better comparability with historical years. Still, in principle, these can be shifted to reflect seasonal fluctuations better. 

Furthermore, the general formulation enables the application of the presented model in other sectors which are exposed to weather uncertainties - like agriculture - or to be used with alternative input time series. In principle, these are not limited to weather data - for example, correlated load time series from different sectors can be used to generate a set of random but consistent load profiles. 

\newpage

\section*{Methods}

\begin{footnotesize}
	
	In the following, the proposed method for the generation of random weather data as well as the central formulation of the power plant dispatch optimization are described by mathematical equations. 
	
	\vspace{\baselineskip}
	
	\textbf{Input Data Structure} 
	\vspace{\baselineskip}
	
	The structure of the input data is described by \eqref{eq:input}. While the rows of the matrix represent different time steps from 1 to \textit{t}$_{\text{max}}$, each column represents a single time series from 1 to \textit{p}$_{\text{max}}$. In the context of weather data, these could either represent different parameters (like temperature, solar radiation, etc.) or different locations.  
		
	\begin{eqnarray}
		\textit{\textbf{X}}^\text{in} = \begin{bmatrix} 
			x_{1,1}^\text{in} & \dots  & x_{1,p_\text{max}}^\text{in} \\
			\vdots & \ddots & \vdots\\
			x_{t_\text{max},1}^\text{in} &  \dots & x_{t_\text{max},p_\text{max}}^\text{in}
		\end{bmatrix} = (x_{t,p}^\text{in}) \in \mathbb{R} ^{t_\text{max} \times p_\text{max}}
		\label{eq:input}
	\end{eqnarray}
	
	\vspace{\baselineskip}
	\textbf{Normalization of Input Data} 
	\vspace{\baselineskip}
	
	In order to normalize the different input data time series, the empirical distribution function is calculated separately for each input time series using \eqref{eq:edf} and \eqref{eq:one}. 
	
	\begin{eqnarray}
		F_p(x)=\frac{1}{t_\text{max}} \sum_{t=1}^{t_\text{max}}\textbf{1}_{[x_{t,p}^\text{in} \leq x]} \quad \forall p \in \{1,...,p_\text{max}\}
		\label{eq:edf}
	\end{eqnarray}
	
	with: 
	
	\begin{eqnarray}
		\textbf{1}_{[x_{t,p}^\text{in} \leq x]} = \begin{cases} 1 & x \leq x_{t,p}^\text{in} \\ 0 & \,x > x_{t,p}^\text{in} \end{cases} \quad \forall t \in \{1,...,t_\text{max}\}, \forall p \in \{1,...,p_\text{max}\}
		\label{eq:one}
	\end{eqnarray}
	
	The time series are then transformed to a standard normal distribution using the inverse standard normal distribution function \eqref{eq:norminv}.
	
	\begin{eqnarray}
		x_{t,p}^\text{norm}=norminv(F_p(x_{t,p}^\text{in}))  \quad \forall t \in \{1,...,t_\text{max}\}, \forall p \in \{1,...,p_\text{max}\}
		\label{eq:norminv}
	\end{eqnarray}
	
	In order to allow the following computation of the DFT for each interval, the structure of the normalized time series is changed to a 3D matrix as shown in \eqref{eq:xnorm1} and \eqref{eq:xnorm2}. In our case the indices $\tau$ represent the hour of the year from 1 to \textit{$\tau$}$_{\text{max}}$, $p$ the time series and $i$ the historical weather year from 1 to \textit{$i$}$_{\text{max}}$.
	
	\begin{eqnarray}
		(\chi^\text{norm}_{\tau,p,i}) \in \mathbb{R} ^{\tau_\text{max} \times p_\text{max} \times i_\text{max}}
		\label{eq:xnorm1}
	\end{eqnarray}
	
	\begin{eqnarray}
		\chi^\text{norm}_{\tau,p,i}=x_{(i-1)\cdot \tau_\text{max}+\tau,p}^\text{norm} \quad \forall i \in \{1,...,i_\text{max}\}, \forall p \in \{1,...,p_\text{max}\}, \forall \tau \in \{1,...,\tau_\text{max}\}
		\label{eq:xnorm2}
	\end{eqnarray}
	
	\vspace{\baselineskip}
	\textbf{Fourier Transformation and Spectral Analysis}
	\vspace{\baselineskip}
	
	Complex Fourier coefficients $c_{k,p,i}$ \eqref{eq:dft_coef} are calculated for each frequency \textit{k}, for each parameter \textit{p} and for each interval \textit{i} of the 3D input data matrix \eqref{eq:dft}.  
	
	\begin{eqnarray}
		(c_{k,p,i}) \in \mathbb{C}^{\tau_{\text{max}} \times p_{\text{max}} \times i_{\text{max}}} 
		\label{eq:dft_coef}
	\end{eqnarray}
	
	\begin{eqnarray}
		c_{k,p,i}=\sum_{j=1}^{\tau_\text{max}} e^{-2\pi i \cdot \frac{(j-1)\cdot (k-1)}{\tau_\text{max}}} \cdot \chi^{\text{norm}}_{j,p,i} \quad \forall k \in \{1,...,\frac{\tau_\text{max}}{2}\}, \forall p \in \{1,...,p_\text{max}\}, \forall i \in \{1,...,i_\text{max}\}
		\label{eq:dft}
	\end{eqnarray}
	
	The resulting correlation information is described by the Matrix $\Sigma_\textit{k} \in \mathbb{R}^{2p_\text{max} \times 2p_\text{max} }$, which contains the covariances of all real and imaginary parts for frequency $k$ among all parameters as described in \eqref{eq:corr_matrix}. Each element of $\Sigma_\textit{k}$ is calculated using \eqref{eq:corr}, considering all combinations of real and imaginary parts accordingly. 
	
	\begin{eqnarray} 
		\Sigma_{k}=\begin{bmatrix} 
			
			\sigma^k_{Re_1,Re_1} & \dots & \sigma^k_{Re_1,Re_{p_\text{max}}} & \sigma^k_{Re_1,Im_1} & \dots & \sigma^k_{Re_1,Im_{p_\text{max}}} \\
			\vdots & \ddots & \vdots & \vdots & \ddots & \vdots \\
			\sigma^k_{Re_{p_\text{max}},Re_1} & \dots & \sigma^k_{Re_{p_\text{max}},Re_{p_\text{max}}} & \sigma^k_{Re_{p_\text{max}},Im_1} & \dots & \sigma^k_{Re_x,Im_{p_\text{max}}} \\
			\sigma^k_{Im_1,Re_1} & \dots & \sigma^k_{Im_{p_\text{max}},Re_{p_\text{max}}} & \sigma^k_{Im_1,Im_1} & \dots & \sigma^k_{Im_1,Im_{p_\text{max}}} \\
			\vdots & \ddots & \vdots & \vdots & \ddots & \vdots \\
			\sigma^k_{Im_{p_\text{max}},Re_1} & \dots & \sigma^k_{Im_{p_\text{max}},Re_{p_\text{max}}} & \sigma^k_{Im_{p_\text{max}},Im_1} & \dots & \sigma^k_{Im_{p_\text{max}},Im_{p_\text{max}}} \label{eq:corr_matrix} \\
			
		\end{bmatrix} \\ \forall k \in \{1,...,\frac{\tau_\text{max}}{2}\} \nonumber		
	\end{eqnarray}
	
	with:
	
	\begin{eqnarray}
		\sigma^k_{Re_a,Re_b}=\frac{1}{i_\text{max}^2} \sum_{i=1}^{{i_\text{max}}}  \sum_{j=1}^{{i_\text{max}}} \frac{1}{2} Re(c_{k,a,i}-c_{k,a,j})\cdot Re(c_{k,b,i}-c_{k,b,j}) \label{eq:corr}
		\\ \quad \forall k \in \{1,...,\frac{\tau_\text{max}}{2}\},\forall a \in \{1,...,p_\text{max}\},\forall b \in \{1,...,p_\text{max}\} \nonumber
	\end{eqnarray}
	
	Additionally, the mean values of the real and imaginary parts are calculated for each frequency according to \eqref{eq:mean_re} and \eqref{eq:mean_im}. The mean values of the real and imaginary parts are then combined in $\mu_{k}$ according to \eqref{eq:mean_total}.
	
	\begin{eqnarray}
		\mu^{Re}_{k,p}=\frac{1}{i_{\text{max}}} \cdot \sum_{i=1}^{i_{\text{max}}} Re(c_{k,p,i})   \forall k \in \{1,...,\frac{\tau_\text{max}}{2}\}, \forall p \in \{1,...,{p_\text{max}}\}
		\label{eq:mean_re}
	\end{eqnarray}
	
	\begin{eqnarray}
		\mu^{Im}_{k,p}=\frac{1}{i_{\text{max}}} \cdot \sum_{i=1}^{i_{\text{max}}} Im(c_{k,p,i})   \forall k \in \{1,...,\frac{\tau_\text{max}}{2}\}, \forall p \in \{1,...,{p_\text{max}}\}
		\label{eq:mean_im}
	\end{eqnarray}

	\begin{eqnarray}
		\mu_{k}=\begin{bmatrix} 
			\mu^{Re}_{k,1} \\
			\vdots \\
			\mu^{Re}_{k,{p_\text{max}}} \\
			\mu^{Im}_{k,1} \\
			\vdots \\
			\mu^{Im}_{k,{p_\text{max}}} \\ 
		\end{bmatrix}
		\quad \forall k \in \{1,...,\frac{\tau_\text{max}}{2}\}
		\label{eq:mean_total}
	\end{eqnarray}
	
	\textbf{Generation of Random Weather Data}
	\vspace{\baselineskip}
	
	Let the random variable ${C}_k \in \mathbb{R}^{2{p_\text{max}}}$ represent the real and imaginary parts of frequency $k$ from all $p_\text{max}$ time series, as shown in \eqref{eq:coef_struct}.
	
	\begin{eqnarray}
		\textbf{C}_k=\begin{bmatrix}
			c^{Re}_{k,1} \\
			\vdots \\
			c^{Re}_{k,{p_\text{max}}} \\
			c^{Im}_{k,1} \\
			\vdots \\
			c^{Im}_{k,{p_\text{max}}} \\       
		\end{bmatrix} \quad \forall k \in \{1,...,\frac{\tau_\text{max}}{2}\}
		\label{eq:coef_struct}
	\end{eqnarray}
	
	The random values are generated based on the covariance matrix $\Sigma_k$ \eqref{eq:corr_matrix} and the corresponding mean values $\mu_k$ \eqref{eq:mean_total} following the probability density function of the $2p_\text{max}$-dimensional multivariate normal distribution as shown in \eqref{eq:mvnrnd}.

	\begin{eqnarray}
		f_{C_k}(\mathbf{x})= \frac{1}{ \sqrt{(2\pi)^{2{p_\text{max}}} \det(\mathbf{\Sigma_k})} } \exp \left(-\frac{1}{2}({\mathbf x}-{\boldsymbol\mu_k})^{\top}{\mathbf{\Sigma_k}}^{-1}({\mathbf x}-{\boldsymbol\mu_k}) \right) \quad \forall k \in \{1,...,\frac{\tau_\text{max}}{2}\}
		\label{eq:mvnrnd}
	\end{eqnarray}
	
	Finally, an inverse DFT is applied to the random coefficients. Each time step $t$ of parameter $p$ is calculated using \eqref{eq:idft}.
	
	\begin{eqnarray}
		&x^{out}_{t,p}=\frac{1}{\tau_{max}} [\sum_{k=1}^{\tau_{max}/2} e^{2\pi \frac{i \cdot (t-1)}{\tau_{max}}} \cdot (c^{Re}_{k,p}+i\cdot c^{Im}_{k,p}) \label{eq:idft} \\
		&+ \sum_{k=\tau_{max}/2}^{\tau_{max}} e^{2\pi \frac{i \cdot (t-1)}{\tau_{max}}} \cdot (c^{Re}_{\tau_{max}-k,p}-i\cdot c^{Im}_{\tau_{max}-k,p})] 
		\quad \forall t \in \{1,...,\tau_\text{max}\}, \forall p \in \{1,...,p_\text{max}\} \nonumber	
	\end{eqnarray}
	
	\textbf{Postprocessing}
	\vspace{\baselineskip}
	
	Finally, the generated time series are denormalised \eqref{eq:denormalize} using the empirical distribution function previously determined for the individual parameters \eqref{eq:edf}. 
	
	\begin{eqnarray}
		x^{gen}_{t,p}=F^{-1}_p(\frac{1}{2}-\frac{1}{\sqrt{\pi}} \int_{ x^{out}_{t,p}}^{\inf} e^{-t^2}dt) \quad \forall t \in \{1,...,\tau_\text{max}\}, \forall p \in \{1,...,p_\text{max}\}
		\label{eq:denormalize}
	\end{eqnarray}
	
	\textbf{Regional Clustering}
	\vspace{\baselineskip}
	
	In order to model the European energy system, hourly resolved weather time series of temperature, solar radiation and wind speed at different altitudes are required for about 4.000 nodes of the transmission grid. Due to the large number of locations and parameters, the application of the model would be impractical, since the variance-covariance matrices representing the dependency of the individual spectral components would become very large.
	
	In order to allow the application in this context, an aggregation method is proposed, which clusters a suitable number of time-series based on their similarity to each other. The aggregation is done by first randomly assigning input time series to clusters. Based on this, a representative time series with the lowest mean square difference to all other time series of the cluster is selected for each cluster. In the next iteration, a new cluster assignment will be carried out, in which each time series will be assigned to the representative time series of a cluster to which it is most similar (with respect to the RMSE).  The process of selecting representative time series per cluster and reassignment is repeated until no more reassignments occur.
	
	In this case study, European historical weather data from the years 1996 to 2016 is used. The data is provided by the DWD as part of the \textit{REA 6} model and is publicly available under \cite{DWD_REA6}. While the DWD provides a bunch of different weather parameters, we use wind speed, temperature and global solar radiation, since these have the highest impact on the behaviour of different parts of the energy system. 
	
	\vspace{\baselineskip}
	\textbf{Electricity Market Model}
	\vspace{\baselineskip}
	
	In the following section, a reduced set of equations for the economic dispatch model is presented. The objective function \eqref{eq:ObjectiveFunctionDispatch_simplified} aims to minimize the variable operating costs of each power plant $k$ in each time step $t$ by considering the product of the current power feed-in $P_{k,t}$ and specific costs $c_k$ which includes the costs for primary fuel, CO\textsubscript{2}-emissions and operation and maintenance. 
	
	\begin{equation} \label{eq:ObjectiveFunctionDispatch_simplified} 
		min \sum_{t \in \mathcal{M}_t}^{} \sum_{k \in \mathcal{M}_k}^{} c_k \cdot P_{k,t}
	\end{equation}
	
	\begin{equation} \label{eq:Constraint_loadCoverage} 
		\sum_{k \in \mathcal{M}_{k,m}}^{} P_{k,t} + \sum_{k \in \mathcal{M}_{ES,m}}^{} P^{\text{ES}}_{k,t} + P^{\text{RES}}_{m,t} = D_{m,t} + P^{\text{NEx}}_{m,t} \quad \forall \: m \in \mathcal{M}_{m} , \forall \: t \in \mathcal{M}_t
	\end{equation}
	
	\eqref{eq:Constraint_loadCoverage} ensures that for each bidding zone $m$, the sum of the electricity demand $D_{m,t}$ and the net export position $P^{\text{NEx}}_{m,t}$ is covered by the sum of feed-in from conventional power plants, energy storages $P^{\text{ES}}_{k,t}$ and RES $P^{\text{RES}}_{m,t}$ in each time step. Note that $P^{\text{ES}}_{k,t}$ can take negative values in this example in the case of a charging process increasing the overall electricity demand in the corresponding hour.
	
	\begin{subequations}  
		\begin{align}
			P^{\text{NEx}}_{m,t} = \sum_{m_2 \in \mathcal{M}_{m}}^{} (P^{\text{Ex}}_{m_1,m_2,t} - P^{\text{Ex}}_{m_2,m_1,t}) \quad \forall \: m_1 \in \mathcal{M}_{m} , \forall \: t \in \mathcal{M}_t \label{eq:Constraint_NExport} \\
			0 \leq P^{\text{Ex}}_{m_1,m_2,t} \leq \text{NTC}_{m_1,m_2} \quad \forall \: (m_1,m_2) \in \mathcal{M}_{m} , \forall \: t \in \mathcal{M}_t \label{eq:Constraint_NTC}
		\end{align}
	\end{subequations}
	
	\eqref{eq:Constraint_NExport} and \eqref{eq:Constraint_NTC} describe the possible power exchange between bidding zones with $(P^{\text{Ex}}_{m_1,m_2,t}$ - $P^{\text{Ex}}_{m_2,m_1,t})$ representing the power balance between two connected bidding zones. The maximum permissible power exchange is limited by the net transfer capacity (NTC) between two bidding zones. Note that the resulting physical flows are based on network topology and network resources and do not necessarily match the determined power balances of each bidding zone. 
	
	\begin{equation} \label{eq:Constraint_P_konv_minMax} 
		0 \leq P_{k,t} \leq P^{\text{max}}_{k} \quad \forall \: k \in \mathcal{M}_{k} , \forall \: t \in \mathcal{M}_t
	\end{equation}
	
	\begin{equation} \label{eq:Constraint_P_RES_minMax} 
		0 \leq P^{\text{RES}}_{m,t} \leq P^{\text{RES,max}}_{m,t} \quad \forall \: m \in \mathcal{M}_{m} , \forall \: t \in \mathcal{M}_t
	\end{equation}
	
	\begin{equation} \label{eq:Constraint_P_ES_minMax} 
		-P^{\text{ES,maxCharge}}_{m,t} \leq  P^{\text{ES}}_{k,t} \leq P^{\text{ES,maxDischarge}}_{m,t} \quad \forall \: k \in \mathcal{M}_{ES} , \forall \: t \in \mathcal{M}_t
	\end{equation}
	
	\eqref{eq:Constraint_P_konv_minMax} limits the current feed-in from conventional power plants to the maximum available capacity and \eqref{eq:Constraint_P_RES_minMax} limits the current feed-in from RES to the maximum available yield in that time step. Generally, due to the very low or non-existing variable cost of RES their contribution to the load coverage is fully integrated. However, in cases of surplus RES or due to other constraints like must-Run restrictions, for example, for the provision of heat from combined heat and power plants, there might be times when the RES cannot be integrated fully into the market. The amount of such \textit{dumped energy} is the difference between the decision variable $P^{\text{RES,max}}_{m,t}$ and the maximum available yield in that time step. 
	
	\eqref{eq:Constraint_P_ES_minMax} limits the charging and discharging process of energy storages and \eqref{eq:Constraint_StorageStateEquation} represents the storage state equation with the two-way efficiency $\eta^{\text{ES}}_k$. \eqref{eq:Constraint_StorageCapacity} keeps the amount of energy inside the storage within the permissible available capacity. The initial amount of energy inside the ES is set with \eqref{eq:Constraint_StorageCapacityIni}.
	
	\begin{subequations}  
		\begin{align}
			E^{\text{ES}}_{k,t} = E^{\text{ES}}_{k,t-1} - P^{\text{ES}}_{k,t} \cdot \eta^{\text{ES}}_k \quad \forall \: k \in \mathcal{M}_{ES} , \forall \: t \in \mathcal{M}_t \textbackslash \{1\} \label{eq:Constraint_StorageStateEquation} \\
			0 \leq E^{\text{ES}}_{k,t} \leq E^{\text{ES,max}}_{k} \quad \forall \: k \in \mathcal{M}_{ES} , \forall \: t \in \mathcal{M}_t \label{eq:Constraint_StorageCapacity} \\
			E^{\text{ES}}_{k,t} = E^{\text{ES,Ini}}_{k} \quad \forall \: k \in \mathcal{M}_{ES} , \forall \: t = 1 \label{eq:Constraint_StorageCapacityIni}
		\end{align}
	\end{subequations}
	
	In order to provide the essence of things, the presented set of equations is limited to the core functionalities of an economic dispatch model, i.e., the load coverage per bidding zone, the representation of possible trade exchange between bidding zones and technical constraints of generation and storage units. The equations are often extended to account for more complex relationships, for example, by introducing start-up and shut-down times as well as a minimum down- and operation times and by considering cooling curves or operating-point-dependent efficiencies of thermal power plants. Further common formulations of the dispatch problem include the consideration of reserve power, a detailed mapping of CHP plants, planned and forced outages and energy storages (especially water storages), formulations for electricity-based loads such as electric vehicles, power-to-heat and power-to-gas devices, flow tariffs and flow-based representations of the limited grid capacities. Some features include the usage of binary variables or lead to non-linearities which strongly increase the complexity of the problem. In our case, the dispatch optimization also considers the operation of flexible loads like EVs and HPs as described in \cite{KROGER2023120406}.  
	
\end{footnotesize}

\section*{Acknowledgements}
This research was partly funded by the \textit{Deutsche Forschungsgemeinschaft} (DFG) within the project \textit{Probabilistische mittel- und langfristige Preisprognosen in Strommärkten}. The authors
would like to further extend their gratitude to Johan Löfberg for developing and publishing the optimization toolbox \textit{YALMIP} \cite{Lofberg2004}. The authors gratefully acknowledge the computing time provided on the Linux HPC cluster at Technical University Dortmund (LiDO3), partially funded in the course of the Large-Scale Equipment Initiative by the German Research Foundation (DFG) as project 271512359.

	\renewcommand*{\bibfont}{\footnotesize}
	\printbibliography[heading=bibintoc]

\end{document}